\RequirePackage{fix-cm}
\documentclass[smallextended]{svjour3}       
\smartqed  
\usepackage{graphicx}
\journalname{Experimental Astronomy}
\begin{document}

\title{Design of a 7m Davies-Cotton Cherenkov telescope mount for the high energy
section of the Cherenkov Telescope Array}

\titlerunning{Design of a 7\,m Cherenkov Telescope for CTA}

\author{A.C. Rovero \and
P. Ringegni \and
G. Vallejo \and
A.D. Supanitsky \and
M. Actis \and
A. Botani \and
I. Ochoa \and
G. Hughes \and
for the CTA Consortium
}


\institute{A.C. Rovero \and A.D. Supanitsky \at
           Instituto de Astronom\'ia y F\'isica del Espacio (CONICET-UBA), Argentina \\
	     Members of ``Carrera del Investigador Cient\'ifico of CONICET''\\
             \email{rovero@iafe.uba.ar}
           \and
           P. Ringegni \and G. Vallejo \and M. Actis \and A. Botani \and I. Ochoa \at
           Unidad de Investigaci\'on y Desarrollo GEMA, Ingenier\'ia Aeron\'autica (UNLP), Argentina\\
	      \email{ringegni@ing.unlp.edu.ar}
           \and
           G. Hughes \at
           Deutsches Elektronen-Synchrotron (DESY), Zeuthen, Germany
}

\date{Received: date / Accepted: date}

\maketitle

\begin{abstract}
The Cherenkov Telescope Array is the next generation ground-based observatory for the study
of very-high-energy gamma-rays. It will provide an order of magnitude more sensitivity and
greater angular resolution than present systems as well as an increased energy range
(20 GeV to 300 TeV). For the high energy portion of this range, a relatively large area
has to be covered by the array. For this, the construction of $\sim$7 m diameter Cherenkov
telescopes is an option under study. We have proposed an innovative design of a Davies-Cotton
mount for such a telescope, within Cherenkov Telescope Array specifications, and evaluated its
mechanical and optical performance. The mount is a reticulated-type structure with steel tubes
and tensioned wires, designed in three main parts to be assembled on site. In this work we show
the structural characteristics of the mount and the optical aberrations at the focal plane for
three options of mirror facet size caused by mount deformations due to wind and gravity.


\end{abstract}

\section{Introduction}
\label{intro}
The present generation of ground-based observatories like HESS, MAGIC and VERITAS have led
the field of very-high-energy gamma-ray astronomy for the last few years, achieving great success
through significantly increasing the rate of source discovery and making detailed observations on many
interesting objects. The international scientific community is now organizing the next generation of
ground-based instruments, the {\it Cherenkov Telescope Array} (CTA). With sensitivity an order of
magnitude better than present systems and improved angular resolution, CTA will observe the full sky
by constructing two observatories, one in each hemisphere, covering an extended energy range
(20 GeV to 300 TeV) (Actis et al. 2011). To span such a large energy range, three different
sizes of Cherenkov telescope will potentially be constructed. Large size telescopes (LSTs) and
medium size telescopes (MSTs) are under study to cover the low and intermediate energy range,
respectively. Small size telescopes (SSTs) will be used to cover a large area (about 10 km$^2$) and
detect the highest energy gamma rays ($>$1 TeV), for which a $\sim$3-4 m diameter mirror is required
as reflecting area for each SST. 

Two types of optical system are being considered for the SST, Schwarzschild-Couder (SC)
(Vassiliev et al. 2007) and Davies-Cotton (DC) (Lewis 1990). The first of these, with primary and
secondary mirrors, has short focal length and hence a reduced plate scale which translates into
a small camera. Two SC designs have been proposed for the SST (White et al. 2011) whose reliability
will be investigated using the prototypes that are currently under construction. The DC design has a
simpler structure and mirrors and is already proven, as it is the type of telescope used by present
experiments. The camera required is, however, larger and potentially more expensive. To balance the
costs of the camera and mount, a first alternative was considered for the DC design, {\it i.e.} to
cover the area of $\approx$10 km$^2$ with fewer SSTs with a relatively large mirror area. Based on
past experience and Monte Carlo simulations, it was specified that this alternative should have a
6-7 m diameter tessellated reflecting surface (dish). The dish would be an array of hexagonal facets,
each with three mounting points, two of them with actuators to align the mirror.
The detector at the focal plane is an array of $\sim$1400 photomultipliers, each of $\sim$0.25$^\circ$
angular diameter, at a focal distance $f \approx$11 m, covering a field of view (f.o.v.) of 8-10$^\circ$.
The Point Spread Function (PSF) for the SST has to be less than the angular size of the camera pixel.
A new type of DC SST is also under consideration using the FACT camera approach (Bretz et al.
2011), which uses cheaper solid photodetectors and thus allows a reasonable cost balanced to be
achieved. This telescope uses a smaller dish of diameter about 4 m.

All the characteristics mentioned above for the SST, and other similar ones for the MST and the LST,
were specified by the CTA Consortium based on previous experience and the science requirements.
Considering these specifications, MC simulations are being performed to determine which is the best
combination of telescope types and parameters for the array that will optimize the science needs.
Meanwhile, any proposed telescope has to fulfill the general characteristics required for the relevant
telescope type.

We have proposed an innovative design of Davies-Cotton mount for the SST of CTA and evaluated its
mechanical and optical performance under both operational and extreme conditions. The first approach 
used a 6 m dish (Actis et al. 2010, Rovero et al. 2011). Following this experience, the decision
was made to enlarge the dish to 7 m diameter and use more regular layouts for the facets, for
which a better optimized mount was obtained. In this work we present the final mechanical design
of a 7 m DC mount and the study of the optical aberrations related to mount deformations for three
different mirror facet sizes and under various telescope operating conditions.

\section{Telescope mount: mechanical design}
\label{sec:1}

\begin{figure*}[!t]
  \centering
  \includegraphics[width=.99\textwidth]{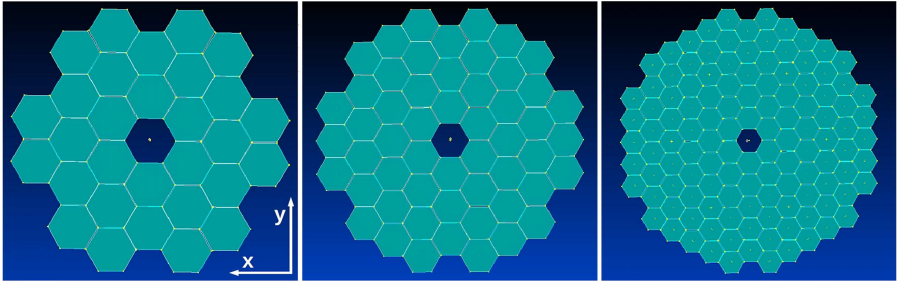}~\hfill
  \caption{Telescope-dish layouts for different mirror facet sizes. {\it Left:} for facet size 120 cm
(from flat to flat edge, {\it i.e.} twice the apothem). The plane $x-y$ of the right-handed coordinate system
is indicated. {\it Centre:} for facet size 90 cm. {\it Right:} for facet size 60 cm.}
   \label{fig.mirrors}
 \end{figure*}

Considering the general specifications for the SST optical system, a pixel size of 50 mm and a focal
length of 11.46 m were adopted, resulting in an $f/d \sim$1.65 and a pixel angular diameter of 0.25$^\circ$.
The focal plane was designed for a camera with maximum f.o.v. of 10$^\circ$. For the construction of a 7 m
dish, hexagonal mirror facets of three sizes ($L$) were considered, $L$ = 60, 90 and 120 cm from flat to flat
({\it i.e.} twice the apothem). Figure \ref{fig.mirrors} shows the dish layouts considered for the mount
design, which are hexagonal arrays of mirrors with the facets at the corners removed to make the dish more
circular and to reduce optical aberrations. The mirrors are mounted on a spherical surface (following the
DC concept) with a spacing of 2 cm to avoid contact between facets. In Table \ref{table.mirrors},
the number of facets and the total reflecting area for each dish layout are shown. Also shown in the table
is an estimation of effective area considering the shadowing of the structure and a 2 m diameter camera.  

\begin{table}[!h]
\caption{Number of mirror facets, total reflecting area, and effective area for the dish layout
of Figure \ref{fig.mirrors}.}
\label{table.mirrors}
\begin{tabular}{llll}
\hline\noalign{\smallskip}
Mirror size $L$ [cm] & Number of facets & Total mirror area [m$^2$] & Effective area [m$^2$] \\
\noalign{\smallskip}\hline\noalign{\smallskip}
120  &  30 & 37.4 & 32.2\\
90   &  54 & 37.9 & 32.1\\
60   & 120 & 37.4 & 31.3\\
\noalign{\smallskip}\hline
\end{tabular}
\end{table}

The main mechanical requirements established by CTA for the SST mount are summarized in Table
\ref{table.mecanica}. To avoid resonance phenomena the eigen frequencies of the structure are
required to be greater than 2.5 Hz. Under emergency wind speed, the telescope has to be able
to drive to the parking position, in which the structure should resist the survival wind speed.

\begin{table}[h]
\caption{Mechanical requirements for the SST.}
\label{table.mecanica}
\begin{tabular}{ll}
\hline\noalign{\smallskip}
Parameter                       & Condition             \\
\noalign{\smallskip}\hline\noalign{\smallskip}
Eigen frequencies               & $>$ 2.5 Hz            \\
Positioning speed, azimuth      & 180 $^\circ$/min      \\
Positioning speed, elevation    & 90 $^\circ$/min       \\
Acceleration, azimuth           & $>$ 1 $^\circ$/sec$^2$   \\
Acceleration, elevation         & $>$ 0.5 $^\circ$/sec$^2$ \\
Tracking precision              & $<$ 6 arcmin         \\
Pointing precision              & $<$ 10 arcsec        \\
Temperature range, operational  & -10 $^\circ$C, 30 $^\circ$C\\
Temperature range, survival     & -15 $^\circ$C, 60 $^\circ$C\\
Wind speed, operational         & $<$ 50 km/h          \\
Wind speed, emergency           & $<$ 100 km/h         \\
Wind speed, survival            & $<$ 180 km/h         \\
\noalign{\smallskip}\hline
\end{tabular}
\end{table}

\begin{figure*}[!h]
  \centering
  \includegraphics[width=.271\textwidth]{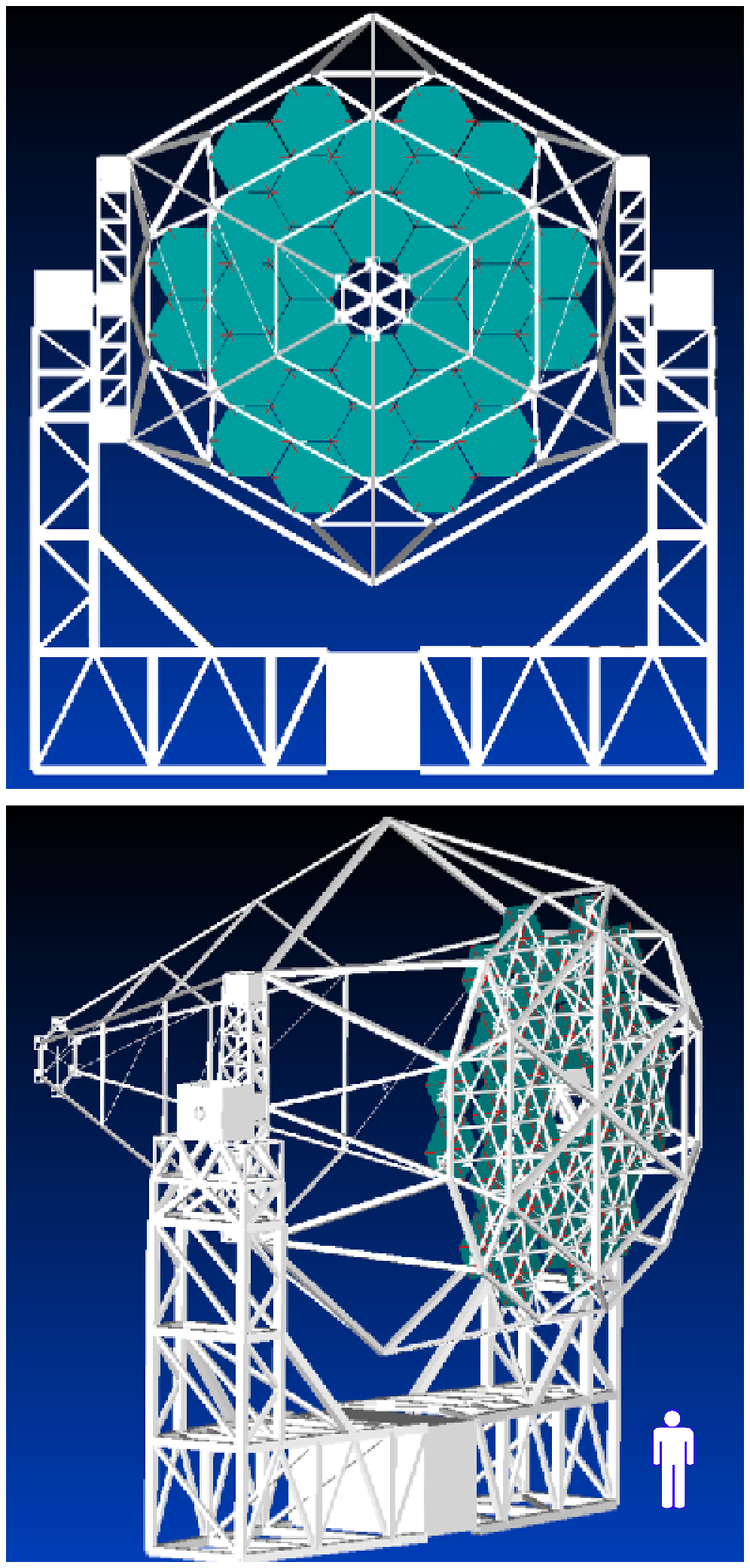}~\hfill
  \includegraphics[width=.729\textwidth]{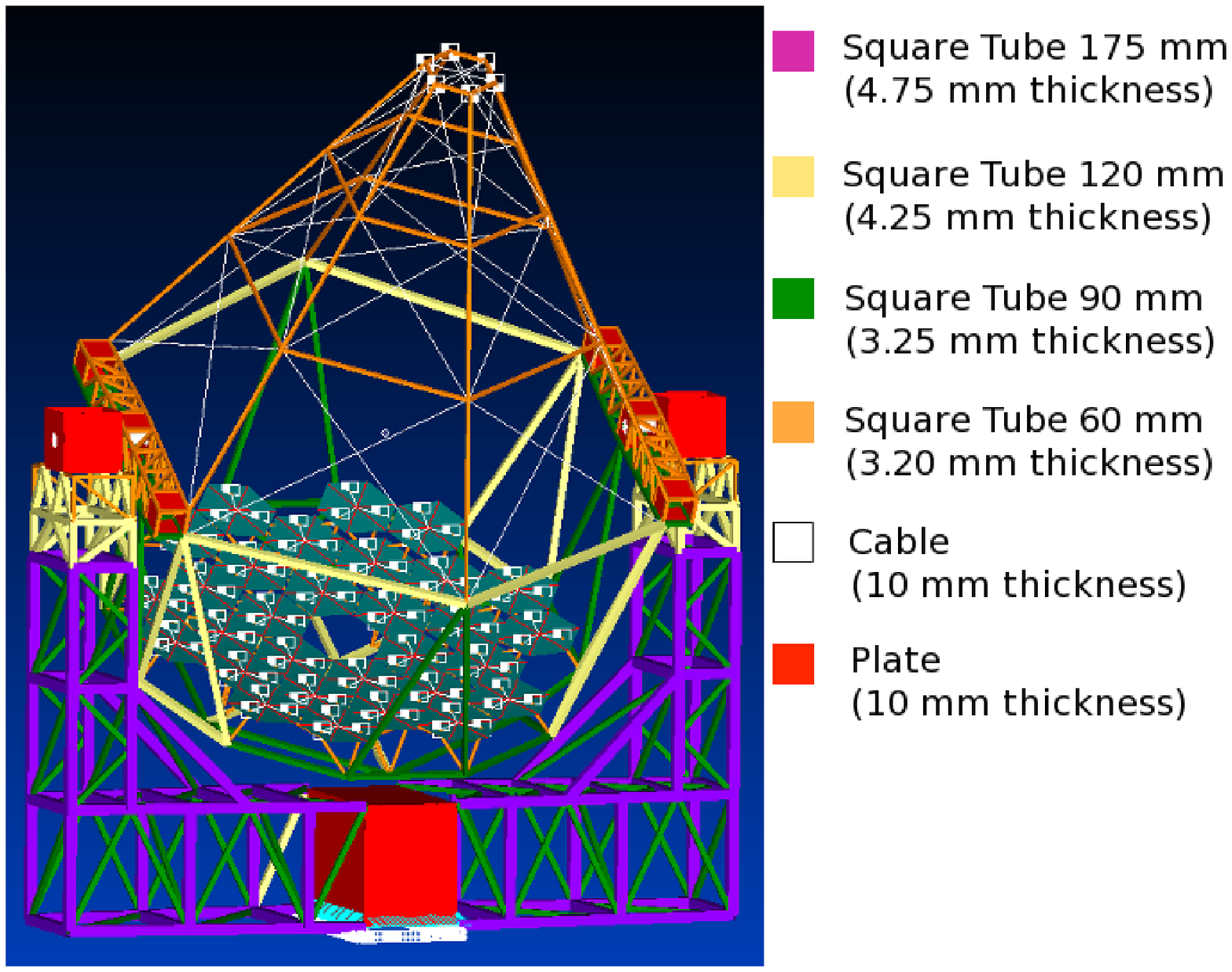}
  \caption{Telescope mount with mirror layout for facets of 120\,cm. {\it Left:} two views of the
structure: from the front ({\it top}), and from the back ({\it bottom}). {\it Right:} Color coded
to differentiate structure components and sizes.}
   \label{fig.struct}
 \end{figure*}

The mount designed is a reticulated-type structure with square cross section tubes of F-24
steel (low carbon) and tensioned wires (see Fig. \ref{fig.struct}). For simplicity, tubes are
reduced to 4 different cross section sizes. The structure consists of three different parts: the
azimuthal or lower structure; the elevation or upper structure; and the mirror support structure.
Each has its own requirements and resulting mechanical solutions. For the azimuthal structure, the fact
that it will be exposed mainly to flectional and torsional forces, determines leads to the adoption of 
a space truss whose main components are welded structural tubes. With ease of transport in mind, the
azimuthal structure is expected to be built in three sections, two vertical and one horizontal truss.
A single central shaft with pinion and crown located at the bottom of this structure provides the
azimuthal movement.
The elevation mount consists of a truncated pyramid with a hexagonal base, in which each face is a truss
with cordons and uprights consisting of structural tubes and tensioners. It becomes complete with two
drawer trusses on the sides to transmit the forces through the shafts to the azimuthal structure.
Finally, the mirror support structure is a two-layer stereo structure of hexagonal shape. It completes
the previous structure and fulfills the dual purpose of supporting the load of the mirrors and wind's
action on them, also providing a stiffening plane in the elevation structure. The mirror support
structure is the only part of the mount to be modified according to the chosen dish layout. Elevation
and azimuth double drive systems were adopted to improve structure movements and better resist wind
moments. Servomotors in all shafts give constant torque at low speed.

Structural analysis, including buckling, was carried out to design the mount and to have good safety
margins. The hypotheses assumed for the Finite Element Model are:
\begin{itemize}
\item Linear-elastic behavior.
\item Isotropic Material, Young's Modulus 2.1$\times$10$^{11}$ Pa, Poisson's Modulus 0.32,
density 7800 kg m$^{-3}$.
\item Truss and Shaft simulated by beam elements.
\item Sheet metal parts (truss-shaft links) simulated by plate elements.
\item Tensioners simulated by rod elements.
\item Mirrors simulated by mass elements, located on the tips of the truss-mirror links.
\item Camera simulated by mass elements, located on 6 points of the camera-truss link structure.
\item For simulations it is considered that shafts are impelled of relative movements with linkages.
\item Camera mass of 1600 kg and a mirror mass of 33 kg m$^{-2}$.
\item Body load of 9.81 m s$^{-2}$ applied to simulate gravity.
\item Wind loads on structure simulated by force per length unit.
\item Wind loads on mirrors simulated by single forces on the tips of mirror linkages.
\item Wind on sheet metal parts simulated by force per unit area (dynamic pressure).
\item Nodes on the bottom of the sheet metal parts of the azimuthal structure with pinned constraints.
\item Total mass of azimuthal structure of 8900 kg (28,984 elements).
\item Total mass of elevation structure of 4400 kg (9154 elements).
\item Total mass of mirror support structure of 1200 kg (3300 elements).
\item Total mass of elevation shafts of 780 kg (40 elements).
\end{itemize}

Simulations were carried out for three cases: normal operation, positioning, and parking. For normal
operation, the tracking ability of the system is tested considering all loads under maximum operational wind
speed (50 km/h). Structure deformations are computed for all possible elevation angles and used to verify
the optical performance (see Section \ref{sec.aberrations}). The positioning case is the critical operational
condition for the structure mechanisms and drives. It simulates the situation of positioning the telescope
from any given point to the parking position under a critical wind speed of 100 km/h. Maximum shaft moments
were found to be 60,000 N\,m for the elevation shaft (30,000 N\,m each) and 42,000 N\,m for the azimuth
shaft. The parking (or survival) case is the critical situation that ensures the structural integrity of
the telescope. It considers winds of 180 km/h and allows the dimensioning of the structural components. The
result of this simulation gave a security margin of 1.97 for maximum stresses, and 1.83 for the elasto-plastic
behavior analysis. A fourth case was considered to check the first natural frequency of the system, which
was estimated to be 3.2 Hz (second natural frequency 4 Hz).

The final step in the design process is to perform a detailed engineering of the mount. After this is
complete, an optimization of the costs for both the prototype and the production telescope mount will be
possible. The CTA Consortium is still performing the studies to decide what the final array would be and,
in particular, what type of SST should be adopted. Once these studies are concluded, an optimization of the
costs will be a necessity.

\section{Optical Aberrations}
\label{sec.aberrations}

Optical aberrations are related to the dispersion of photons collected at the focal plane.
As a measure of such dispersion, the diameter of a circle containing 80 \% of the collected light
at the focal plane is considered (herein denoted D80). To test the optical aberrations of the telescope
mount, no consideration is given to the intrinsic aberrations of mirror facets, {\it i.e.} the roughness
and mirror deformations are not taken into account. We have performed the analysis of optical aberrations
for our design in two steps, firstly studying system aberrations for the dish layouts proposed in the
previous section, and secondly we have considered the contribution of mechanical deformations as obtained
in Section \ref{sec:1} from simulations under normal operation.

\subsection{System aberrations}
The intrinsic aberration of several wide-field optical systems available for IACTs have been studied
previously ({\it e.g.} Schliesser \& Mirzoyan 2005). DC optical systems are known to have relatively 
low aberrations for large off-axis incidence angles, which depend on the layout of the facets for the
dish under consideration. Here we compute the optical aberrations specifically for the mirror layouts
proposed for our DC design and by computing D80 instead of the commonly used PSF ({\it i.e.} the RMS
of the photon distribution at the focal plane for each axis). This also allows us to estimate the
aberrations caused by structure deformations by comparison with intrinsic system aberrations
(Section \ref{mech.deform}).

The arrays of hexagonal facets (Figure \ref{fig.mirrors}) are mounted on a spherical
structure with radius of curvature $R$ = 11.46 m, {\it i.e.} equal to the focal length. Each facet
is a spherical mirror with radius 2$R$, following the DC design (Lewis 1990). In our coordinate system
the center of the dish is placed at the point (0,0,-R), {\it i.e.} the origin is at the center of
the focal plane (camera), $z$ is the optical axis, $x$ is parallel to the Earth's surface, and $y$
is perpendicular to both, as shown in Figure \ref{fig.mirrors}.
The direction of any incident photon is defined by two angles relatively to the $z$-axis: $\phi_x$ in the
$x-z$ plane (azimuth) and $\phi_y$ in the $y-z$ plane (elevation). We describe the reflected photon
using the angles $\xi$ and $\eta$. Thus, the incident and reflected photons are characterized by
$(\phi_x ,\phi_y)$ and $(\xi ,\eta)$, respectively.

 \begin{figure}[!h]
  \centering
  \includegraphics[width=.46\textwidth]{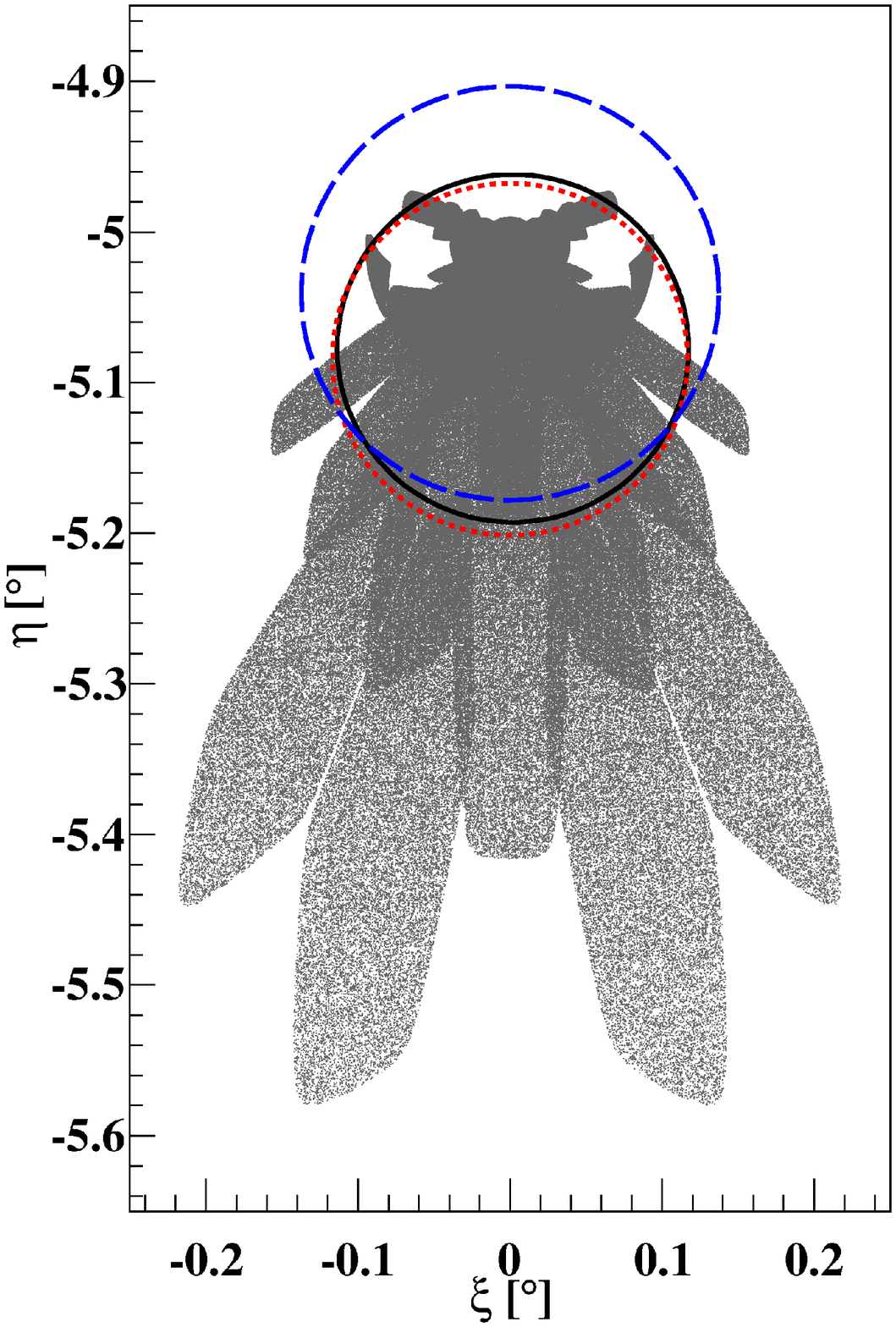}
  \includegraphics[angle=90, width=.3365\textwidth]{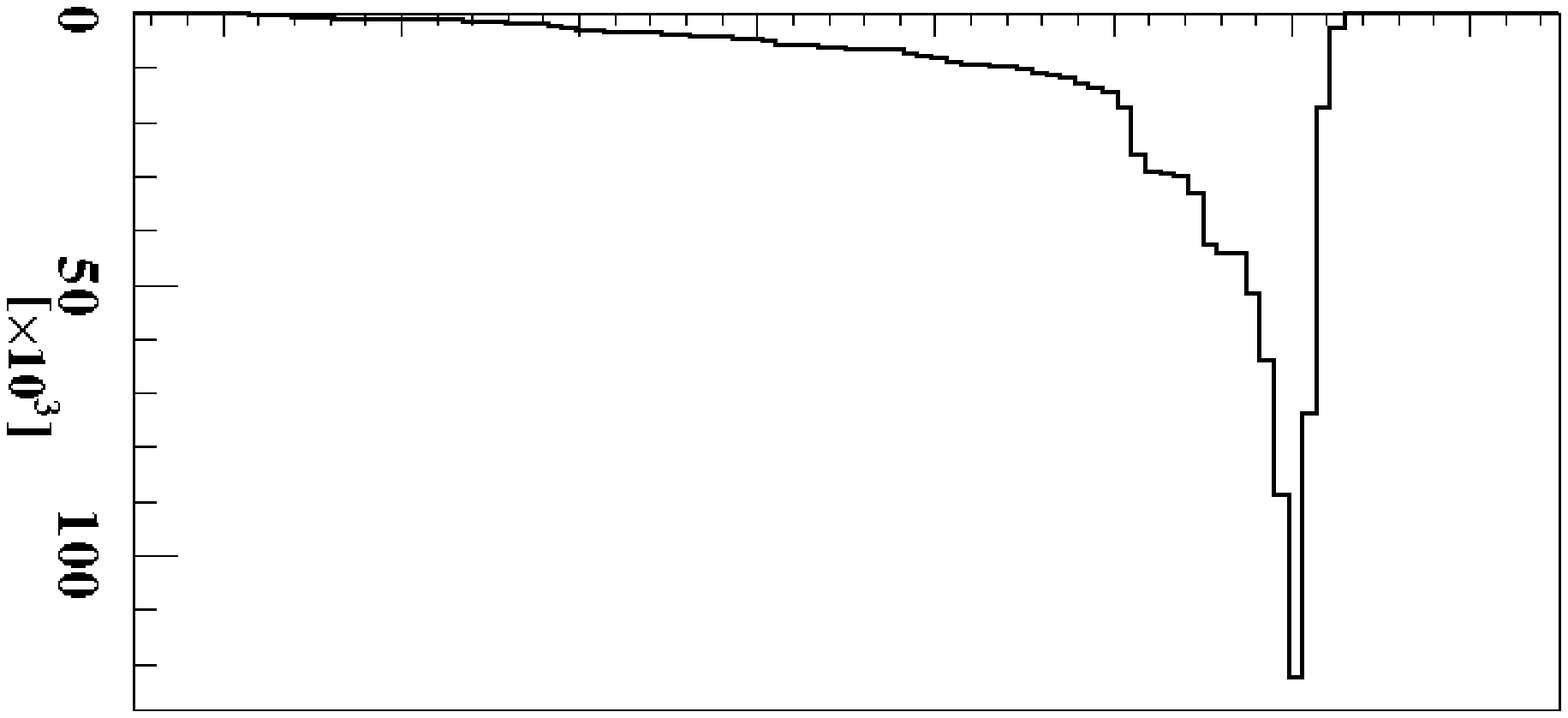}
  \caption{Simulated distribution of collected photons at the focal plane using the dish formed of 30
120 cm facets (see Figure \ref{fig.mirrors}), for an angle of incidence of 5$^\circ$ on the
elevation axis. {\it Left:} bidimensional distribution with D80 computed for a circle centered at the
median (blue-dashed), mean (red-doted) or optimized (black-solid) position. {\it Right:} projection on the
elevation axis.}
  \label{fig.focalPlane}
 \end{figure}

Given the incidence of a parallel photon beam on the dish, a non-pointlike distribution of reflected
photons is formed at the focal plane due to system aberrations. To characterize this distribution we
used our own MC ray-tracing code for 10$^6$ photons homogeneously distributed over the dish. An example of
such a distribution at the focal plane is shown in Figure \ref{fig.focalPlane}, for the dish formed by 30
120 cm facet mirrors, and for a rather extreme angle of incidence, $(\phi_x,\phi_y) = (0^\circ,5^\circ)$.
In the same figure, the distribution on the elevation axis is shown, indicating the strong concentration
of reflected photons at the specular position ($\eta = -5^\circ$). The shape of the distribution is
typical for a DC telescope.

The definition of D80 given above (Section \ref{sec.aberrations}) is not unique. The most intuitive
interpretation would be to take
the diameter of a circle centered at $(<\!\xi\!> ,<\!\eta\!>)$ that encloses 80 \% of the photons
($D_{80}^{\mu}$), or to take the median as the circle center ($D_{80}^{med}$). The smallest possible
values of D80 are obtained by choosing the position of the circle in the focal plane that minimizes
its diameter ($D_{80}$). Figure \ref{fig.focalPlane} shows these cases for the particular distribution
shown in the figure. We have obtained D80 using these three definitions for incidence angles on different
planes and shown elsewhere (Rovero et al. 2011) that $D_{80}$ is slightly better than $D_{80}^{\mu}$,
although the latter could be taken as a very good approximation. In this work we use $D_{80}$.

 \begin{figure}[!h]
  \centering
  \includegraphics[width=.9\textwidth]{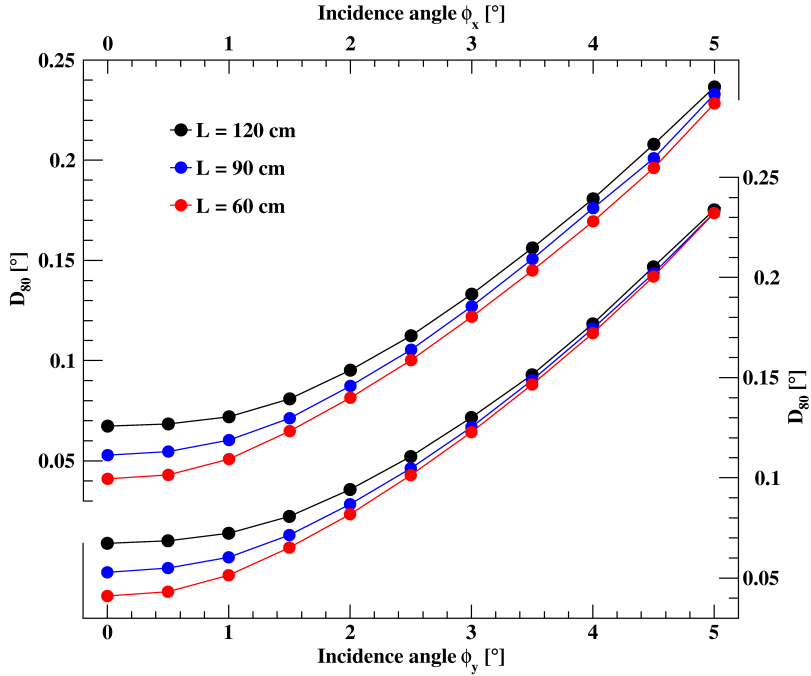}
  \caption{System aberrations ($D_{80}$) as a function of incidence angle for different dish layouts
characterized by the facet size ($L$). Curves with coordinate axis in the upper-left corner correspond to
incidence angles in azimuth, $\phi_x$. Curves with coordinate axis in the lower-right corner correspond to
incidence angles in elevation, $\phi_y$. Uncertainties in the estimation of $D_{80}$ are too small to
be noted.}
  \label{fig.d80facets}
 \end{figure}

Figure \ref{fig.d80facets} shows the results of the simulations, where two sets of curves are presented,
corresponding to incidence angles in the azimuth ($\phi_x$) and the elevation ($\phi_y$) planes. The
small differences in the behavior of the curves reflect the fact that the dishes are not perfectly symmetrical.
The curves indicate that reducing the facet size improves the general performance, as expected, although
for angles $\phi_y \!> \!3^\circ$ the system aberrations are similar for incidence in the elevation plane.
Uncertainties in the estimation of $D_{80}$ should, in principle, be computed by evaluating the minimization
procedure used to find $D_{80}$ in each  case.
Values of $D_{80}$ including uncertainties are always below the pixel size
(0.25$^\circ$) even for the maximum proposed field of view (10$^\circ$) and, consequently, within
specifications.

\subsection{Including mechanical deformations}
\label{mech.deform}
Telescope mount deformations cause mirror facets to change their orientation and,
consequently, to change the photon distribution at the focal plane. As only structural deformations
are considered here, the procedure used was to evaluate the displacements of the positions of the facet
mounting points relative to their ideal positions using the simulation code outlined in Section
\ref{sec:1}. Once the new displaced positions of the mirror facets were found, ray-tracing
was used to find the new $D_{80}$ including mechanical deformations.

\begin{figure}[!h]
  \centering
  \includegraphics[width=0.8\textwidth]{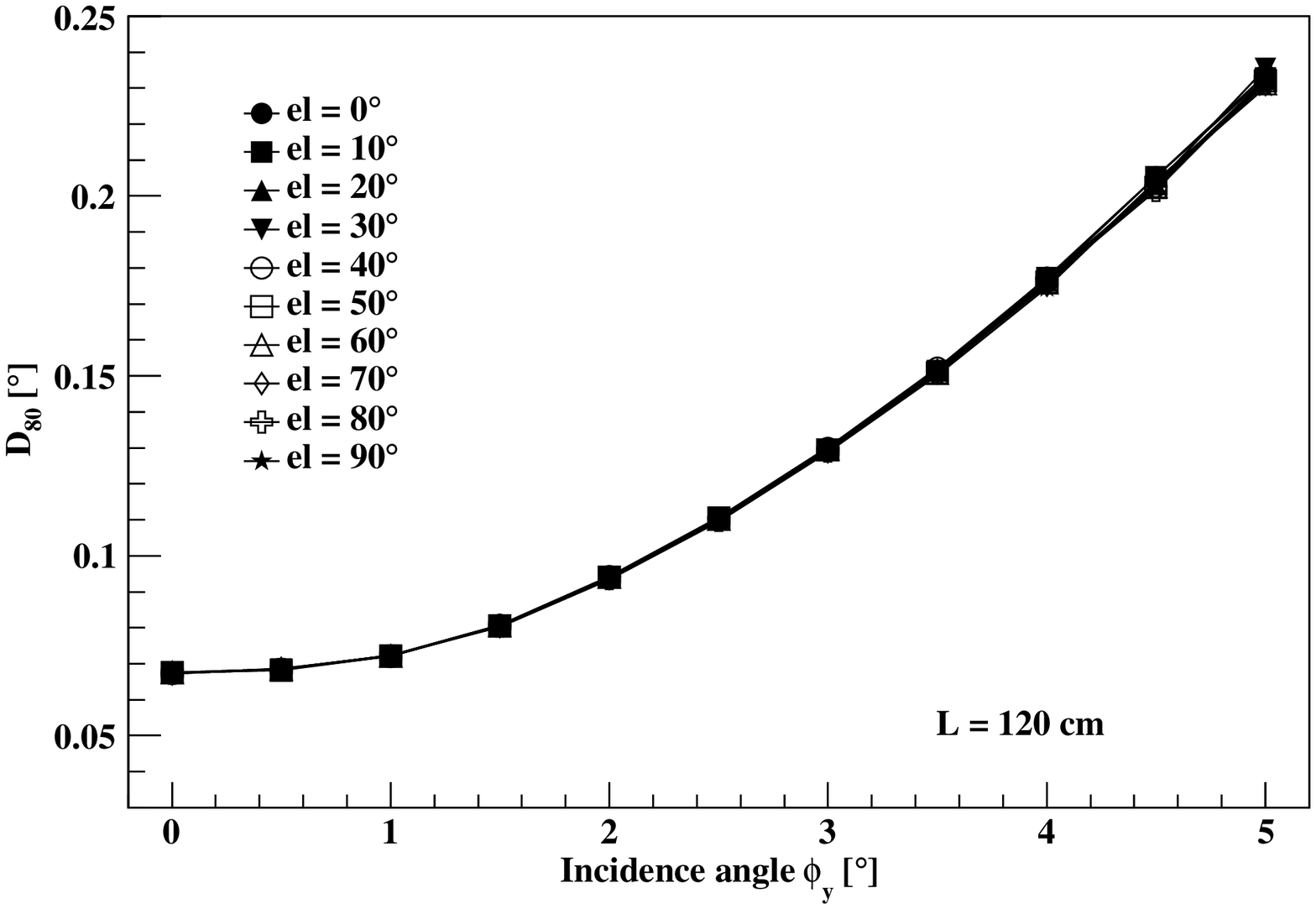}
  \caption{Example of $D_{80}$ as a function of incidence angle, $\phi_y$, at different telescope elevations,
considering structural deformations, shown for the mirror layout corresponding to facets of 120
cm, and under the worst operational conditions. Note that most of the points are overlapping.}
  \label{fig.deform}
 \end{figure}

\begin{figure}[!h]
  \centering
  \includegraphics[width=0.48\textwidth]{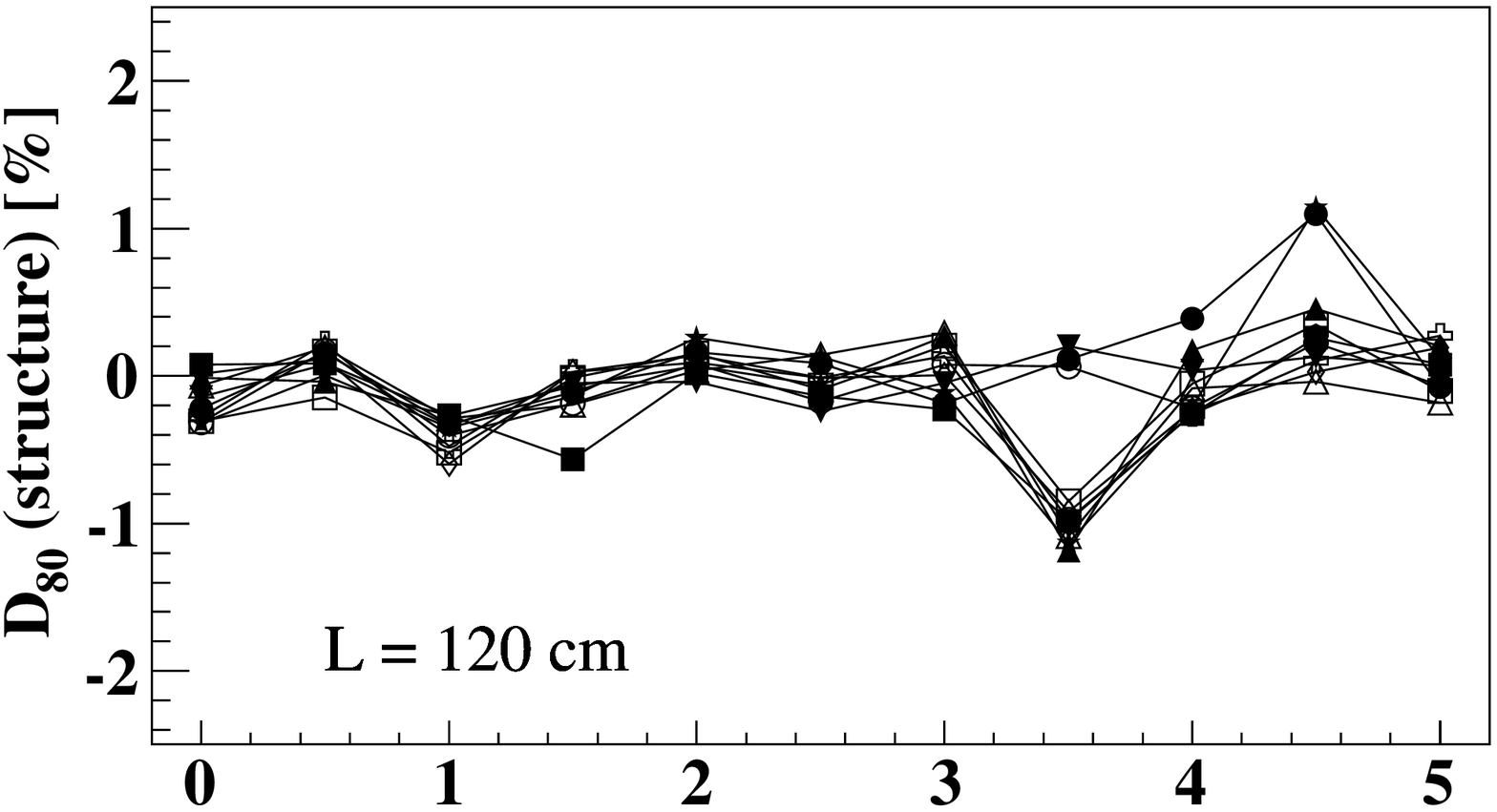}
  \includegraphics[width=0.48\textwidth]{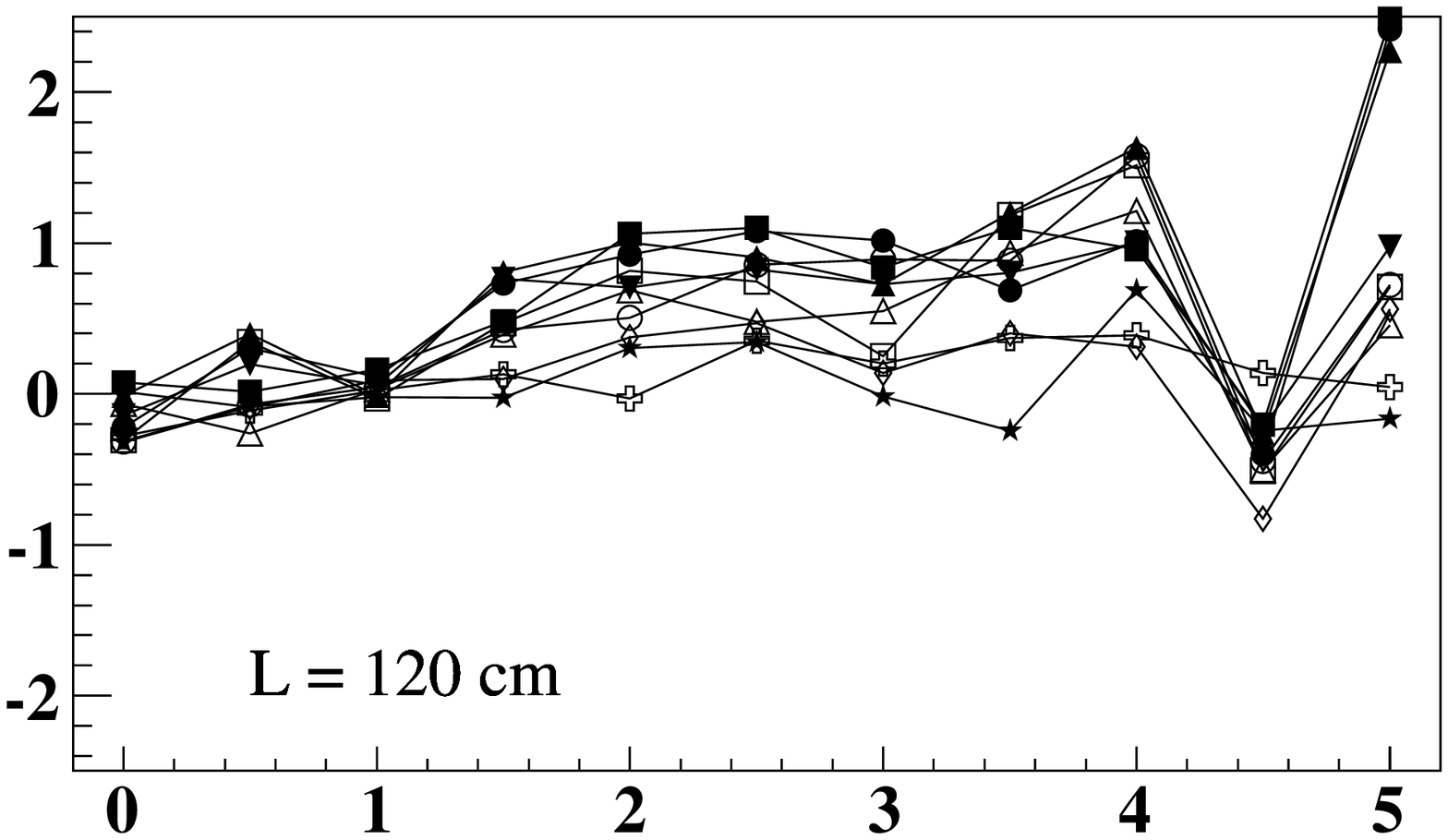}
  \includegraphics[width=0.48\textwidth]{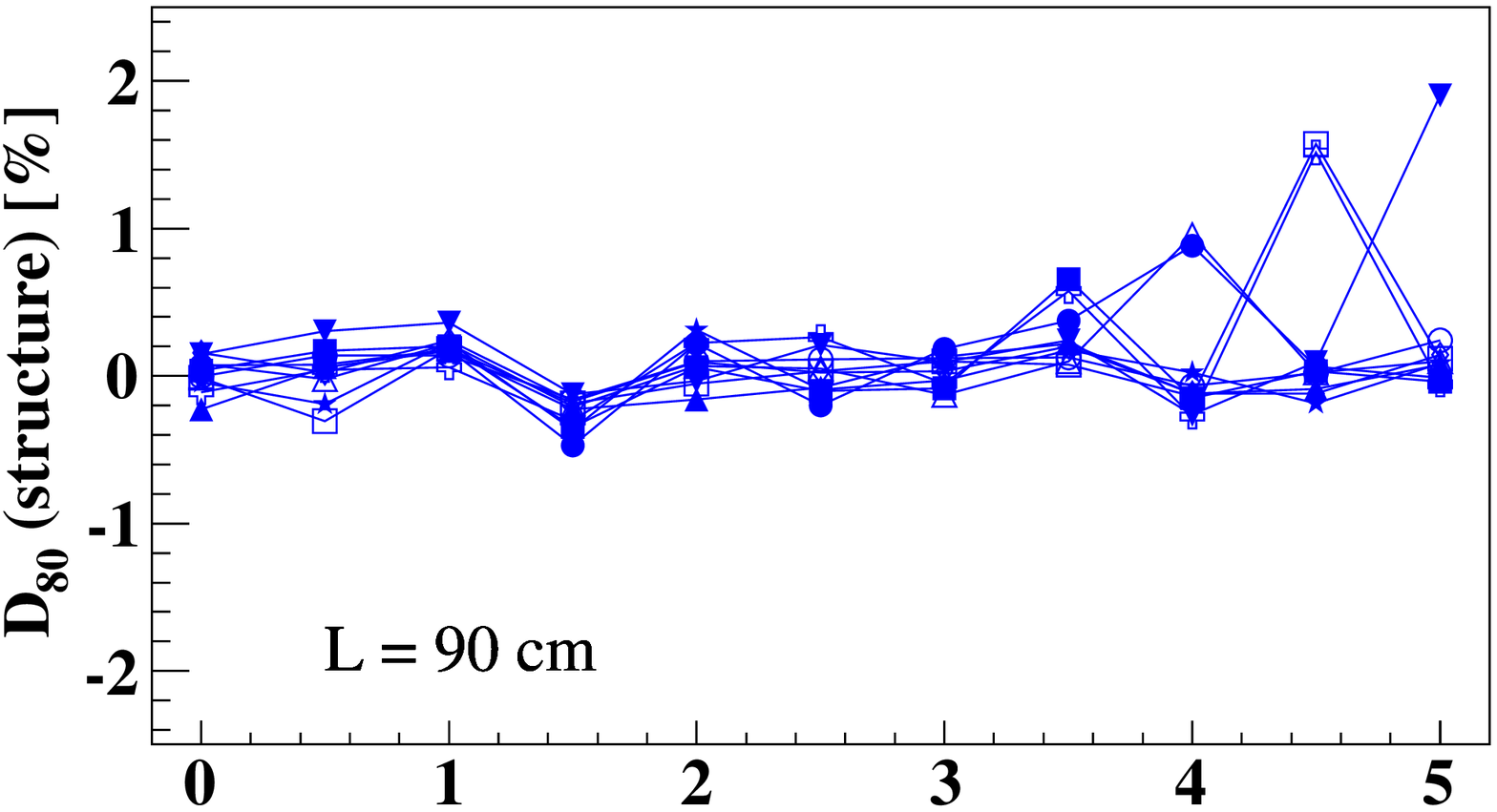}
  \includegraphics[width=0.48\textwidth]{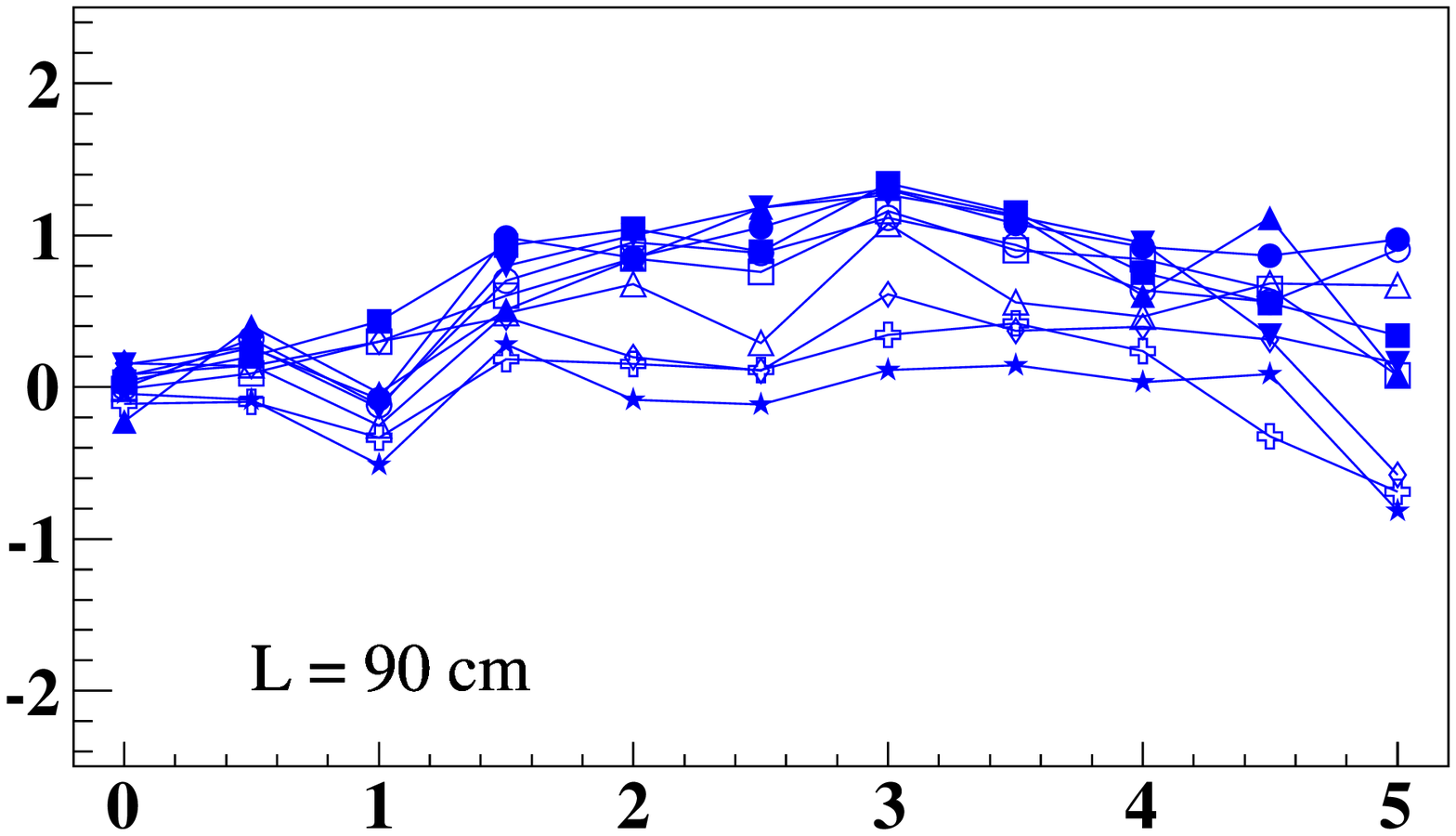}
  \includegraphics[width=0.48\textwidth]{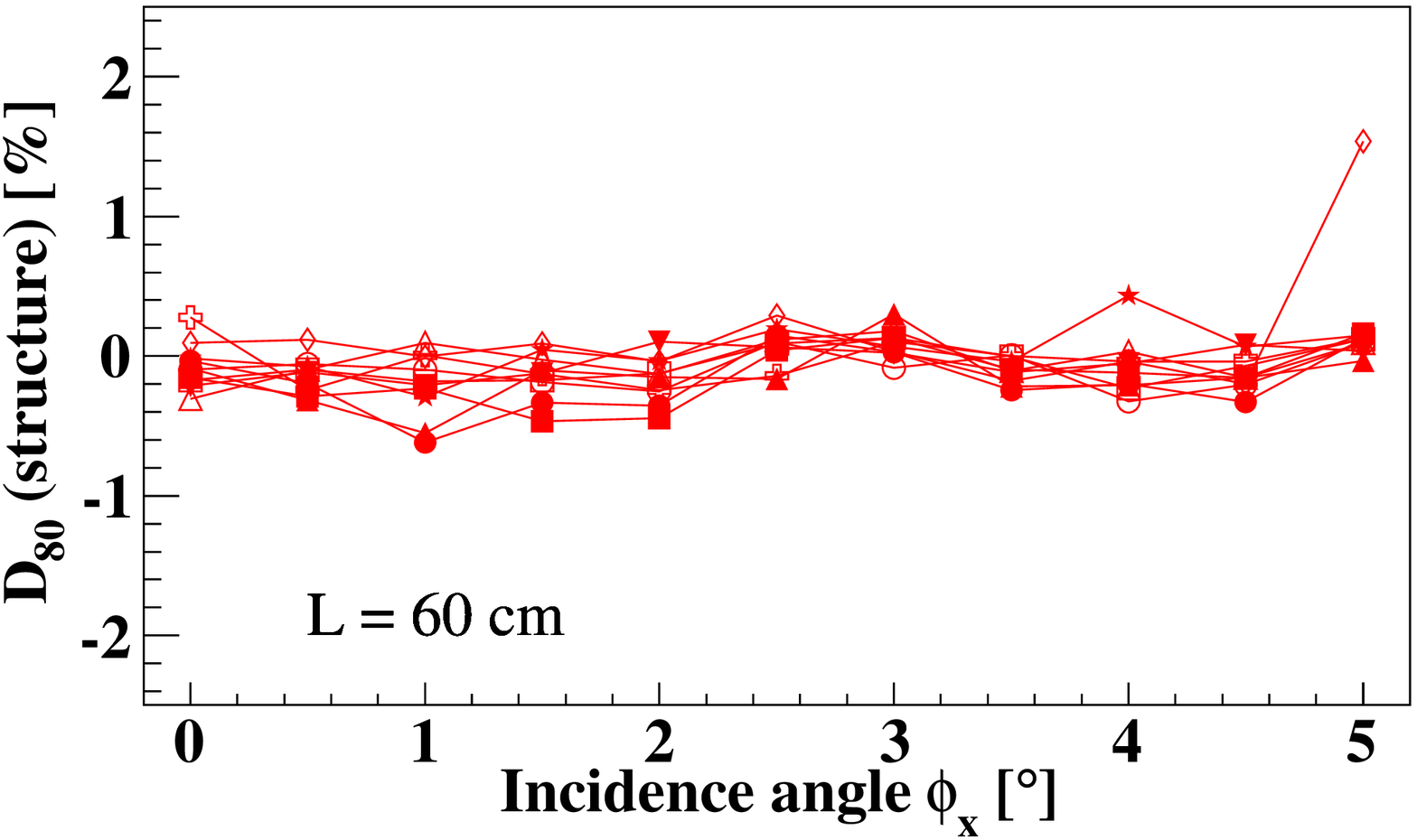}
  \includegraphics[width=0.48\textwidth]{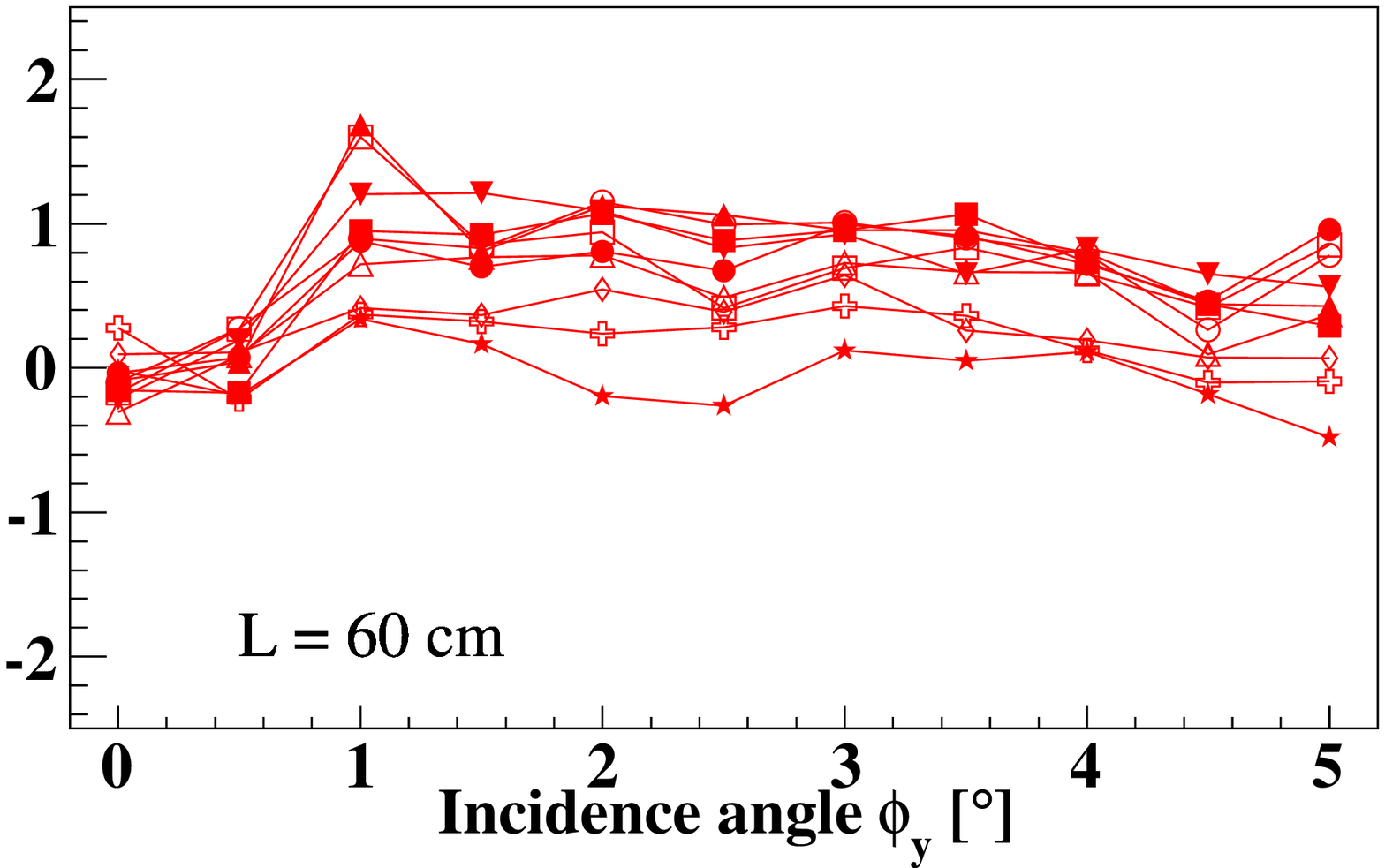}
  \caption{Contribution of structural deformations to optical aberrations, as a percentage of $D_{80}$ Vs.
incidence angle, under the  worst operational conditions. Each figure has ten curves corresponding
to telescope elevations from 0$^\circ$ to 90$^\circ$ using the nomenclature of Figure \ref{fig.deform}.
{\it Left:} azimuth incidence angles, $\phi_x$. {\it Right:} elevation incidence angles, $\phi_y$.
{\it Top:} for $L$ = 120 cm. {\it Middle:} for $L$ = 90 cm. {\it Bottom:} for $L$ = 60 cm.}
  \label{fig.diff}
 \end{figure}

We have calculated the optical aberration in the worst possible case for all three dish layouts of
Figure \ref{fig.mirrors}, by estimating $D_{80}$ including structural deformations as described above.
The worst possible case means under the most extreme wind speed conditions considered for telescope operation,
{\it i.e.} 50 km/h, hitting from the most disadvantageous angle on the structure. We have done this
for both photon incidence angles, $\phi_x$ and $\phi_y$. Gravity causes structural deformations to depend
on elevation angle, so we have computed the system performance as a function of telescope elevation.
As an example, Figure \ref{fig.deform} shows $D_{80}$ under these conditions for elevations 0$^\circ$ to
90$^\circ$, in steps of 10$^\circ$. No significant degradation of the optical performance is apparent in
the example due to mechanical deformations for our mount design. To better appreciate the influence of
structural deformations, Figure \ref{fig.diff} shows the differences in percentage of $D_{80}$ compared to the
aberrations without mechanical deformations. The figure shows all cases under consideration: $\phi_x$ and
$\phi_y$ for $L$ = 60, 90, and 120 cm.

Even for the worst case of telescope operation, Figure \ref{fig.diff} shows that elevation angle
is not relevant for the telescope's optical performance. Within the large field of view under consideration
for the SST (10$^\circ$), the deterioration of the optical performance due to mechanical deformations
remains typically within 1\,\%, with an apparent improvement for small incidence angles, being the largest
deviation 2.5\,\%.
Then, the overall telescope optical aberrations are, within few percent, the system aberrations as shown
in Figure \ref{fig.d80facets}.

\section{Summary}
We have proposed a design for a 7 m Davies-Cotton mount for the Small Size Telescope of CTA, and have
evaluated how optical performance is affected by mount deformations for three mirror facet sizes. The
mount designed is a reticulated-type structure with square cross-section steel tubes and tensioned
wires. The structure consists of three different parts, the azimuthal, the elevation, and the mirror
support structures. Assuming linear-elastic behavior, a Finite Element Model was used to design the mount
components and to evaluate safety margins. Simulations were performed for three cases: normal operation,
positioning, and parking.
For normal operation (wind speed 50 km/h) the structure deformations were computed for all possible
elevation angles and used to evaluate the optical performance. The positioning case (wind speed 100 km/h)
simulates the situation of driving the telescope from any point to the parking position which allowed
the design of the shafts and drives. The parking (or survival) position ensures the structural integrity
of the telescope under the worst conditions (wind speed 180 km/h). The result of this simulation gave a
security margin of 1.97 for maximum stresses, and 1.83 for the elasto-plastic behavior analysis. The
first natural frequency of the system was estimated to be 3.2\,Hz. 

Optical aberrations were computed using our own MC ray-tracing to evaluate the system aberrations and
the contribution to the aberrations caused by mechanical deformations as a function of elevation angle
for three dish layouts using hexagonal mirror facets of 120 cm, 90 cm, and 60 cm. As a measure
of optical aberrations we evaluated the diameter of the smallest circle containing 80 \% of the
photons collected by the system at the focal plane. Our results show that aberrations are lower for
smaller mirror facets, in particular for small incidence angles, although the differences are not significant.
Mechanical deformations do not significantly change the optical performance. Variations in the aberrations
due to this effect are within 1\,\% for most incidence angles, with an apparent improvement for small
angles, being the largest deviation 2.5\,\%. Even considering this extreme case, the aberrations
estimated for our design are always smaller than the pixel size (0.25$^\circ$) and, within few percent,
well described by the system aberrations (without considering mechanical deformations). This conclusion
applies for the worst possible case of telescope operation and for any telescope elevation angle. It
suggests that our mount design might be more rigid than necessary. Considering that the telescope cost
has yet to be studied for our design, one could imagine relaxing the mechanical tolerances to make the
structure cheaper, while still keeping the deformations within specifications. However, the current rigidity
of the structure is in response to the mechanical requirements outlined in Section \ref{sec:1}. Of particular
importance is the survival condition, where the security margin is slightly smaller than a factor two.
While this margin is reasonable, it prevents a significant relaxation of the tolerances.

\begin{acknowledgements}
We thank to the anonymous Referee for very helpful comments. We gratefully acknowledge support
from the agencies and organisations listed in this page: http://www.cta-observatory.org/?q=node/22.
\end{acknowledgements}

\end{document}